\documentclass[fleqn,10pt]{wlscirep}

\usepackage{graphicx,epsfig}
\usepackage{color}
\usepackage{amsmath,amssymb,mathrsfs,amsthm}
\usepackage{amsfonts}
\usepackage{multicol,multirow}
\usepackage{graphicx}
\usepackage{color}
\usepackage{xcolor}
\usepackage{extarrows}
\usepackage{pdfcolmk}
\usepackage{float}
\usepackage[nice]{nicefrac}
\usepackage{amsthm}
\usepackage{latexsym}
\usepackage{dsfont}
\usepackage{setspace}
\usepackage{hyperref}
\usepackage[square,sort,comma,numbers]{natbib}
\usepackage[fit]{truncate}
\usepackage[font=small,labelfont=bf,labelsep=space]{caption}
\usepackage{algorithm}
\usepackage[noend]{algpseudocode}
\usepackage{caption}
\usepackage{enumitem}
\usepackage{framed}
\usepackage{authblk}
\usepackage{lineno}
\usepackage{fancyhdr,lipsum}
\algrenewcommand{\algorithmicrequire}{\textbf{Input:}}
\algrenewcommand{\algorithmicensure}{\textbf{Output:}}
\makeatletter

\hypersetup{
	colorlinks   = true,
	citecolor    = blue
}

\date{}

\def\@citex[#1]#2{\if@filesw\immediate\write\@auxout{\string\citation{#2}}\fi
  \def\@citea{}\@cite{\@for\@citeb:=#2\do
    {\@citea\def\@citea{,\linebreak[0]\hskip0pt plus .2em}%
      \@ifundefined{b@\@citeb}%
    {{\bf ?}\@warning{Citation `\@citeb' on page \thepage\space undefined}}%
      \hbox{\csname b@\@citeb\endcsname}}}{#1}}

\newtheorem{theorem}{Theorem}[section]

\newtheorem{definition}[theorem]{Definition}

\newtheorem{rule-def}[theorem]{Rule}

\numberwithin{equation}{section} \makeatletter
\newsavebox{\savepar}

\title{Isogeny-based Post-Quantum Proxy Signature for Internet of Things}
\author[1]{Somnath Kumar}
\author[2]{Kunal Dey}

\author[3]{Vikas Srivastava}
\author[4]{Sumit Kumar Debnath}
\author[5,6,*]{Ashok Kumar Das}
\author[7,8,*]{Shehzad Ashraf Chaudhry}

\affil[1]{``Department of Mathematics, National Institute of Technology Jamshedpur, Jamshedpur, India'' (e-mail: somnath1691997@gmail.com)}
\affil[2]{``Department of Computer Science and Engineering, SRM University-AP, India'' (e-mail: kunal.d@srmap.edu.in, kunaldey3@gmail.com)}

\affil[3]{``Department of Mathematics, National Institute of Technology, Warangal, India'' (e-mail: vsv@nitw.ac.in)}
\affil[4]{``Department of Mathematics, National Institute of Technology, Jamshedpur 831 014, India'' (e-mail: sd.iitkgp@gmail.com, sdebnath.math@nitjsr.ac.in)}
\affil[5]{``Center for Security, Theory and Algorithmic Research, International Institute of Information Technology, Hyderabad 500 032, India'' (e-mail: iitkgp.akdas@gmail.com, ashok.das@iiit.ac.in)}
\affil[6]{``Department of Computer Science and Engineering, College of Informatics, Korea University, 145 Anam-ro, Seongbuk-gu, Seoul 02841, South Korea''}
\affil[7]{``Department of Computer Science and Information Technology, College of Engineering, Abu Dhabi University, Abu Dhabi, UAE'' (e-mail: ashraf.shehzad.ch@gmail.com)}
\affil[8]{``Department of Software Engineering, Faculty of Engineering and Architecture, Nisantasi University, Istanbul 34398, Turkey''}

\affil[*]{Corresponding authors: ashok.das@iiit.ac.in; ashraf.shehzad.ch@gmail.com}

\begin{abstract}
 The rapid growth of the Internet of Things (IoT) introduces challenges in secure authentication and delegation due to the limited computational capabilities of devices. Proxy signature schemes offer an effective solution by enabling controlled delegation of signing rights to more capable entities, such as gateway nodes. However, most existing schemes rely on classical assumptions that are likely to be broken by quantum adversaries. In this work, we address these challenges by proposing an isogeny-based post-quantum proxy signature scheme, \textit{CSI-PS}. The scheme leverages the hardness of the Group Action Inverse Problem (GAIP) to ensure quantum-resistant security while maintaining efficiency suitable for resource-constrained environments. We further demonstrate its applicability in IoT architectures through a gateway-based delegation model. Our analysis shows that the proposed scheme strikes an effective balance between security and efficiency in terms of computation and communication overhead, along with provable security under the EUF-CMA notion. 
 \vspace{.1cm}\\

\noindent{\small{\bf Keywords:}  Isogeny-based public-key cryptography, IoT, proxy signature, post-quantum cryptography, elliptic curve cryptography.}
\end{abstract}

\begin{document}

\flushbottom
\maketitle
%
%
\thispagestyle{empty}

\section*{Introduction}
In the modern world, machines and devices are increasingly interconnected through the Internet. This concept is known as the Internet of Things (IoT)\cite{gokhale2018introduction}. IoT describes a system of interconnected physical devices that are embedded with sensors and communication technologies, enabling them to gather and share data over the Internet with minimal or no human involvement~\cite{abdul2015internet}. Because of this capability, such devices are often referred to as smart devices. IoT technology has been widely adopted in various applications, including smart homes, healthcare systems, and smart cities~\cite{atzori2010internet,li2015internet}. The rapid evolution of IoT has been further driven by advancements in wireless communication technologies, cloud computing, and edge computing, enabling seamless connectivity among heterogeneous devices. Modern IoT systems often operate in large-scale distributed environments where numerous devices continuously sense, process, and transmit data in real time. This capability allows for automation, operational efficiency, and intelligent decision-making.  In smart home environments, IoT devices facilitate automated control of lighting, temperature, and security systems, thereby improving user comfort and energy efficiency. In healthcare applications, wearable sensors and remote monitoring devices enable continuous tracking of patient health, supporting early diagnosis and timely medical intervention. Similarly, in smart city infrastructures, IoT technologies are utilized for traffic management, waste management, and environmental monitoring, contributing to sustainable urban development.

In IoT environments, data is typically transmitted over open and untrusted networks, which introduces significant concerns related to privacy and security~\cite{yang2017survey}. If such systems are not properly secured, sensitive information may be intercepted, modified, or misused by adversaries. To mitigate these risks, cryptographic schemes are necessary for enabling secure interaction between IoT devices. In particular, digital signatures are widely adopted to guarantee authentication, non-repudiation, and data integrity. They enable the receiver to confirm that the data comes from a genuine source and remains unchanged while being transmitted. However, classical cryptographic methods, including those relying on RSA and the discrete logarithm problem, become insecure in a quantum computing setting because of algorithms like Shor’s algorithm.
~\cite{shor1999polynomial}. This has led to growing interest in post-quantum cryptographic solutions that can resist quantum attacks~\cite{debnath2021delegating}. Among these approaches, isogeny-based cryptography has gained great focus because of its relatively small key sizes and strong security assumptions~\cite{jao2011towards}. These features make it a promising candidate for deployment in settings where resources are restricted, such as IoT deployments. In practical IoT scenarios, most devices, including sensors and embedded nodes, operate under strict limitations in terms of computation, memory, and energy. As a result, performing complex cryptographic operations directly on these devices can negatively impact system performance and battery life. Moreover, in large-scale IoT deployments involving a massive number of devices, continuously generating digital signatures at each node can lead to significant computational overhead. Therefore, designing efficient authentication mechanisms that reduce the computational burden on resource-constrained devices while preserving strong security guarantees remains an important research challenge.

A natural approach to address this challenge is to enable controlled delegation of signing permissions. Proxy signature schemes provide an effective cryptographic primitive to achieve this goal, enabling an original signer to securely and verifiably transfer signing authority to a more capable proxy entity. In the IoT setting, this paradigm aligns well with hierarchical architectures where resource-rich gateway nodes act as intermediaries. In such a framework, an IoT authority can serve as the original signer and authorize a gateway node to sign on its behalf~\cite{mambo1996proxy}. Sensor devices collect data and transmit it to the gateway, which generates proxy signatures in place of the authority before forwarding the authenticated data to the cloud for verification and processing. This delegation significantly reduces the computational overhead on end devices while preserving essential security properties such as authenticity, integrity, and non-repudiation.

Despite their practical advantages, most existing proxy signature constructions depend on classical problems, namely integer factorization and discrete logarithms ~\cite{gokhale2018introduction, yu2012provably}. These assumptions are known to be susceptible to quantum adversaries because of algorithms such as Shor’s algorithm, thereby threatening the long-term security of such schemes. This has led to growing interest in designing post-quantum proxy signature schemes based on quantum-resistant primitives. In recent years, several efforts have explored lattice-based \cite{jiang2010lattice} and hash-based \cite{chandrasekhar2010efficient} approaches for constructing proxy signatures that achieve quantum resistance while maintaining efficiency. However, existing post-quantum proxy signature schemes are still limited either in terms of efficiency, delegation flexibility, or suitability for IoT-specific architectures.
Motivated by the need to explore alternative post-quantum paradigms, investigating new PQC-based proxy signature constructions remains an interesting and important research direction. In particular, isogeny-based cryptography stands out for its small key sizes and solid security guarantees, which motivate the construction of our isogeny-based proxy signature scheme.

In contrast, our work focuses on the practical design and evaluation of an isogeny-based proxy signature scheme. We provide detailed parameter instantiation, explicit size analysis, and a comprehensive study of computational costs. Furthermore, while our construction is general, we demonstrate its applicability in IoT environments, where efficient delegation and reduced computational overhead are critical. Thus, our work complements existing isogeny-based proxy signature constructions by bridging the gap between theoretical design and practical deployment.

\subsection*{Our Contributions}

\begin{figure}
\centering
\includegraphics[scale = 0.5]{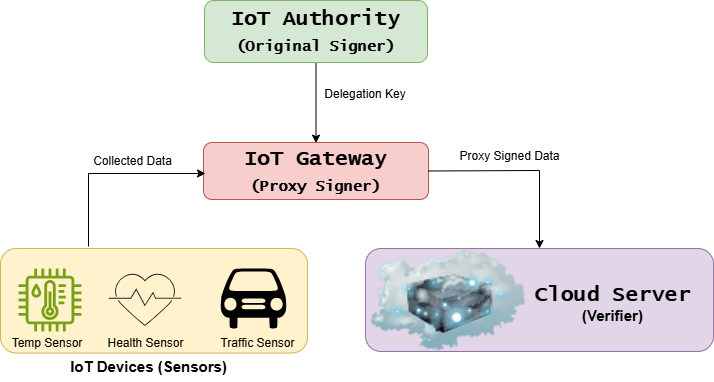}
\caption{Network architecture of the proposed isogeny-based proxy signature scheme for IoT environments.}
\label{Fig_1}
\end{figure}

The main ideas and key contributions of this work are outlined below:

\begin{itemize}

\item We study a practical authentication and delegation setting tailored for Internet of Things (IoT) environments, where resource-constrained devices cannot efficiently perform costly cryptographic operations. In such scenarios, a gateway node can be used to handle signing tasks on behalf of these devices. In our model, an authority securely delegates its signing capability to the gateway, allowing it to generate valid signatures for data produced by IoT sensors. This approach not only ensures authenticated data transmission but also significantly reduces the computational overhead on end devices. To support this setting, we design a proxy signature which is a post-quantum scheme and is built on isogeny-based schemes, which we refer to as \textit{CSI-PS} (Commutative Supersingular Isogeny Proxy Signature). The scheme takes advantage of the small key sizes and strong security properties of isogeny-based constructions, makes it appropriate for constrained IoT deployments. A general view of the system architecture is provided in Figure~\ref{Fig_1}.

\item The proposed \textit{CSI-PS} scheme derives its security from the assumed hardness of the group action inverse problem ($\sf{GAIP}$). We argue that within the random oracle model, it is computationally difficult to construct a valid signature forgery even with adaptive chosen-message access. Furthermore, the scheme meets the fundamental requirements expected from a proxy signature system, including verifiability, identifiability, and undeniability.

\item We carry out a thorough efficiency analysis of \textit{CSI-PS}, including compact size characterization, concrete parameter instantiation, and computational cost evaluation. Specifically, we derive explicit expressions for key and signature sizes, where the Both the original and proxy signers possess public keys of size $L_{0}\lceil \log(p) \rceil$ bits, and their corresponding secret keys are $L_{0}\lceil n \log(2I_{0}+1) \rceil$ bits in length. The proxy share and proxy signature sizes are given by $M_{1}M_2\lceil \log(L_{1}+1) \rceil + M_{1}M_2\lceil n \log(2I_{1}+1) \rceil$ bits.

We further identify practical parameter choices; for instance, under $(16, 8, 8, 2, 65535, 255)$, the scheme achieves a proxy signature size of 3052 bytes, along with a complete instantiation demonstrating the delegation and signing workflow for IoT environments. Additionally, we evaluate the computational cost of the scheme, showing that class group action dominates the complexity, and provide detailed estimates of field multiplications and overall time complexity (see Table~\ref{Compuntationtion}).

\end{itemize}

\subsection*{Outline of the Paper}
The layout of the paper is presented in the subsequent sections. The preliminaries required for the proposed scheme are introduced in Section \ref{Preliminaries}. In Section~\ref{Proposed}, we present the construction of \textit{CSI-PS}. Section~\ref{Security} provides the security analysis. In Section~\ref{Sec: efficiency}, we discuss the communication and computational efficiency, along with a concrete example demonstrating the applicability of \textit{CSI-PS} in IoT environments. Section \ref{Conclusion} provides the concluding remarks.

\section*{Preliminaries}\label{Preliminaries}

\textbf{Notations}. For a finite set $X$, its cardinality is denoted by $|X|$ The symbol $\mathbb{F}_q$ refers to a finite field containing $q$ elements, with $q$ equal to a power of a prime $p$. Closure of a field $\mathbb{F}_q$ is denoted by $\overline{\mathbb{F}}_q$. $[c,d]$ denotes the set of integers $z$ such that $c\le z\le d$.
\begin{definition}{\textbf{Negligible function}}:\\
A function $\epsilon:\mathbb{N} \longrightarrow \mathbb{R}$ is called a negligible function, if for every polynomial $poly(x)$, there exists an integer $N > 0$ such that $\epsilon(x)$ is bounded above by $\frac{1}{poly(x)}$ for all $x > N$.
\end{definition}
\subsection*{Elliptic Curves and Isogenies~\cite{de2017mathematics}}

An elliptic curve over a field $\mathbb{F}_q$ is a smooth projective curve of genus $1$ having at least one $\mathbb{F}_q$-rational point. For char$(\mathbb{F}_q)\not=2,3$, the short Weierstrass affine equation for an elliptic curve over $\mathbb{F}_q$ is $y^{2}=x^{3}+cx+d$, where the point at infinity is $(0:1:0)$, and $c,d\in \mathbb{F}_q$ with $4c^{3}+27d^{2}\not=0$. Throughout the paper, elliptic curves are assumed to be defined over $\mathbb{F}_p$. 
Hasse’s theorem~\cite{silverman2009arithmetic} bounds the number of points on $E(\mathbb{F}_p)$. In particular, there exists an integer $t$ such that  $|E(\mathbb{F}_p)| = p + 1 - t$, with $|t| \leq 2\sqrt{p}$.  \par

\noindent An isogeny $\chi:E_{1}\rightarrow E_{2}$ is an onto morphism inducing a group homomorphism from $E_{1}(\overline{\mathbb{F}}_q)$ to $E_{2}(\overline{\mathbb{F}_q})$. This relation implies that $E_1$ and $E_2$ belong to the same isogeny class. Indeed, $E_1$ and $E_2$ having the equal cardinality assure an isogeny between them. An isogeny sends the identity element $\Theta$ of $E_{1}$ to identity element $\Theta^\prime$ of $E_{2}$. An endomorphism is a morphism from $E$ to itself that fixes the distinguished point $\Theta$. Let $\mathrm{End}(E)$ represent the collection of all map from $E$ to $E$, i.e., endomorphisms. This collection naturally carries a ring structure, where addition and composition define its operations. The Frobenius endomorphism is an endomorphism on $E$ over $\mathbb{F}_q$ and it is defined as $\pi_{E}: (x:y:z) \longmapsto (x^{q}:y^{q}:z^{q})$. Multiplication by $n$ also provides an example of an endomorphism and it is defined as $[n]: Q\longmapsto [n]Q$ and $Q\in E$. Its kernel is known as the $n$-torsion subgroup and this subgroup is represented by $E[n]$.\par
\noindent The standard form of isogeny $\chi$ over $\mathbb{F}_q$ is given by $\chi(x,y)=\left(\frac{\alpha(x)}{\beta(x)}, \frac{\gamma(x)}{\delta(x)}y\right)$, where $\alpha, \beta, \gamma, \delta \in \mathbb{F}_q[x]$, $\alpha$ is co-prime to $\beta$ and $\gamma$ is co-prime to $\delta$ in $\mathbb{F}_q[x]$. We denote the degree of $\chi$ by $deg(\chi)$, where $deg(\chi)=  max\{deg(\alpha), deg(\beta)\}$. If $\left(\frac{\alpha(x)}{\beta(x)}\right)^\prime \neq 0$ then $\chi$ is called separable; otherwise, it is called inseparable. Note that the Frobenius endomorphism over $\mathbb{F}_p$ is an inseparable isogeny. The endomorphism $[m]$ for an elliptic curve $E/\mathbb{F}_p$ is separable precisely when $m $ and $p$ are co-prime. For the elliptic curve $E: y^{2}=x^{3}+cx+d$, the $j$-invariant is given by $j(E)= 1728\frac{4c^3}{4c^3+27d^{2}}$. $E(\overline{\mathbb{F}_q})$ are considered isomorphic when they have identical $j$-invariants. The $p$-torsion subgroup $E[p]$ reflects the nature of the curve: it may either be a cyclic group with $p$ elements or reduce to the trivial group. The former case corresponds to an ordinary curve, while the latter corresponds to a supersingular one.
In characteristic $p$, the $j$-invariant of a supersingular elliptic curve is always contained in the field 
$\mathbb{F}_{p^2}$.

\subsection*{Endomorphism Algebra and Ideal Class Group~\cite{de2017mathematics}}
The endomorphism ring $\text{End}(E)$ is not just a $\mathbb{Z}$-module but also a $\mathbb{Z}$-algebra. The endomorphism algebra of $E$ is denoted by $\text{End}^0(E) := \text{End}(E) \otimes_{\mathbb{Z}} \mathbb{Q}$, making $\text{End}^0(E)$ a $\mathbb{Q}$-algebra.he ring $\mathrm{End}(E)$ is a free $\mathbb{Z}$-module of rank $r$, which leads to the following possibilities. When $r=1$, the algebra $\mathrm{End}^0(E)$ is isomorphic to $\mathbb{Q}$. For $r=2$, it corresponds to an imaginary quadratic field, while in the case $r=4$, it becomes isomorphic to a quaternion algebra. If $E$ is defined over a finite field, the case $r = 1$ cannot occur.
\noindent An \emph{order} $\vartheta$ in a $\mathbb{Q}$-algebra $K$ refers to a subring of $K$ that has the same rank as $K$. For any number field $K$, the ring of integers $\vartheta_K$ is the unique maximal order. In imaginary quadratic number fields, orders are defined by $\vartheta = \mathbb{Z} + f\vartheta_K$, where $f$ denotes an integer known as the conductor of $\vartheta$.

\noindent The Frobenius endomorphism $\pi$ for an elliptic curve $E /\mathbb{F}_q$ satisfies the equation $\pi^{2}-t\pi+q=0$ with $|t|\leq2\sqrt{q}$. Here $t$ is called trace of $\pi$ and $\Delta_{\pi}=t^{2}-4q$ is discriminant of $\pi$. Therefore, the discriminant $\Delta_{\pi}=t^{2}-4q$ of $\pi$ is negative and we have $\pi\in\mathbb{Q}(\sqrt{\Delta_{\pi}})$ which is an imaginary quadratic field.Let $\vartheta$ be an order in a number field $K$. Given a $\vartheta$-ideal $L$, a fractional $\vartheta$-ideal can be defined as
$L' = \omega L = { \omega \alpha \mid \alpha \in L }$, where $\omega \in K^{*}$. A fractional $\vartheta$-ideal $L'$ is said to be invertible if there exists another fractional $\vartheta$-ideal $L''$ such that $L' L'' = \vartheta$. The set of all invertible fractional $\vartheta$-ideals forms a group under ideal multiplication. In particular, for two fractional $\vartheta$-ideals $L' = \omega L$ and $N' = \omega' N$, their product is given by
$L' N' = \omega \omega' , L N$. Let $I$ represent the group of invertible fractional $\vartheta$-ideals and $P$ denotes the subgroup of $I$ consisting of principal fractional $\vartheta$-ideals. The ideal class group is defined as the quotient of the group of fractional ideals by the subgroup of principal ideals, denoted by $cl(\vartheta)=\frac{I}{P}$. Consider a supersingular elliptic curve $E$ over the finite field $\mathbb{F}_p$ The ring of endomorphisms of $E$ that are defined over $\mathbb{F}_p$, written as $\mathrm{End}_p(E)$, can be identified with an order in an imaginary quadratic field. Let $E_p$ represent the set of isomorphism classes of elliptic curves over $\mathbb{F}_p$ for which the corresponding endomorphism ring is isomorphic to $\vartheta$. In this setting, the class group $cl(\vartheta)$ acts on $E_p$ in such a way that the action is both free and transitive.\\
\begin{theorem}
Consider a supersingular elliptic curve $E$ over the finite field $\mathbb{F}_p$, equipped with the Frobenius endomorphism $\pi$, where $p > 3$ and $p \equiv 3 \ (\mathrm{mod}\ 4)$. The necessary and sufficient condition for $\text{End}_p(E)=\mathbb{Z}[\pi]$ is that there exists a unique $e\in\mathbb{F}_p$ so that $E$ is $\mathbb{F}_p$ isomorphic to a Montgomery curve: $y^2=x^3+ex^2+x$, where $\mathbb{Z}[\pi]$ is a subring of $\mathbb{Q}(\sqrt{-p})$. 
\label{Montgomery}
\end{theorem}
\algnewcommand\algorithmicforeach{\textbf{for each}}
\algdef{S}[FOR]{ForEach}[1]{\algorithmicforeach\ #1\ \algorithmicdo}
\begin{algorithm} \caption{The class-group action on a Montgomery curve}
\begin{algorithmic}[1]
\Require A Montgomery curve $E/\mathbb{F}_p$ given by $y^2 = x^3 + e x^2 + x$, with $p = 4l_1 \cdots l_n - 1$, and an integer tuple $(a_1, \ldots, a_n)$.
\While {some $a_i \neq 0$}
\State Sample $x\in \mathbb{F}_p$.
\State Set $w\longleftarrow 1$ if $x^3+ex^2+x$ is a square in $\mathbb{F}_p$,        else set $w\longleftarrow -1$.
\State Suppose $W=\{i : a_i \neq 0, sign(a_i) = w\}$.
\If{$W=\emptyset$}
\State Start from step 2.
\EndIf
\State Set $f\longleftarrow \prod_{i\in W} l_i $.
\State Evaluate $Q \longleftarrow [(p+1)/f]P$, where $P$ is a point in $E$ of order $l_i$.
\ForEach{$i\in W$}
\State Calculate $R\longleftarrow [f/l_i]Q$.
\If{$R=\infty$}
\State skip
\EndIf
\State Compute the isogeny $\phi: E \longrightarrow E^{\prime}$ with $E^{\prime}:y^2=x^3+Bx^2+x$ and $ker(\phi)=R$.
\State Set $Q\longleftarrow \phi(Q)$, $f\longleftarrow f/l_i$, $a_i = a_i - w$
\EndFor
\EndWhile
\Ensure $B$.
\end{algorithmic}
\label{alg:Classgroup}
\end{algorithm}

\subsection*{Commutative Supersingular Isogeny Diffie-Hellman (CSIDH)~\cite{castryck2018csidh}}
\label{CSIDH}
If $E$ is a supersingular curve over $\mathbb{F}_p$ then its ring of $\mathbb{F}_p$-rational endomorphism is an imaginary quadratic order $\vartheta\subseteq\mathbb{Q}(\sqrt{-p})$ which is denoted by $\text{End}_p(E)$. In \textit{CSIDH}~\cite{castryck2018csidh}, we fix $E_0:y^{2}=x^{3}+x$ over $\mathbb{F}_{p}$ for the prime $p=4l_{1}\cdots l_{n}-1$, where $l_1,\cdots, l_n$ pairwise distinct odd primes of relatively small size. The congruence $p\equiv(-1)(\bmod~4)$ ensures that the elliptic curve $E_0$ is supersingular. Since the cardinality of $E_0(\mathbb{F}_p)$ equals $p+1$, it follows that $l_i|(p+1)$. As a consequence, the ideal generated by $l_i$ in $\vartheta$  splits as a product of two conjugate ideals, given by $l_{i}\vartheta=J_{i}\overline{J_{i}}$, where the ideals are defined by   $J_{i}=(l_{i}, \pi-1)$ and $\overline{J_{i}}=(l_{i}, \pi+1)$. Now we have to assume that $J_i$'s are well spread across the class group and it is expected that that an ideal is of the form $J_{1}^{a_{1}} \cdots J_{n}^{a_{n}}$ with $(a_{1},\cdots a_{n})$ randomly chosen from $[-s,s]^n$ for some natural number $s$. Thereby, we may represent the ideal $J_{1}^{a_{1}} \cdots J_{n}^{a_{n}}$ as $(a_{1}, \cdots, a_{n})$. It has been observed that it is sufficient to choose $m$ such that $2s+1\geq\sqrt[n]{|cl(\vartheta)}$. As the trace of Frobenius endomorphism is zero for supersingular elliptic curve over $\mathbb{F}_{p}$, its characteristic equation is $\pi^{2}+p=0$ yielding $\pi=\sqrt{-p}$. Thus, $\text{End}_p(E_{0})=\mathbb{Z}[\sqrt{-p}]=\mathbb{Z}[{\pi}]$. In other words, $E_0$ is isomorphic over $\mathbb{F}_p$ to a Montgomery curve $E_A$ given by
$y^2 = x^3 + e x^2 + x$,
and such a curve can be described solely by the parameter $e$ as stated in Theorem~\ref{Montgomery}.\\
\indent To generate the key, Dora randomly chooses a n-tuple secret key $\textbf{a}=(a_1, \cdots, a_n)$ randomly form $[-s,s]^n$ and computes $[\textbf{a}]=[J_{1}^{a_{1}} \cdots J_{n}^{a_{n}}] \in \mathbb{Z}[\pi]$. She then computes the group action $[\textbf{a}]E_{0}$ using Algorithm \ref{alg:Classgroup}, one may express the given curve in Montgomery form $E_A$ through an isomorphism. In this setting, the parameter $f$ uniquely specifies the curve and is treated as the public key by Theorem \ref{Montgomery}. Here, Dora's secret key - public key pair is $([\textbf{a}], f)$. Similarly Smith chooses his key pairs as $([\textbf{b}], g)$.Next, Dora computes $[\textbf{a}]E_{B}$ by applying 
$[\textbf{a}]$ to $[\textbf{b}]E_{0}$, while Smith similarly evaluates 
$[\textbf{b}]E_{A}$ by acting with $[\textbf{b}]$ on $[\textbf{a}]E_{0}$.
Ultimately, both participants arrive at a common key given by $[\textbf{a}][\textbf{b}]E_{0} = [\textbf{b}][\textbf{a}]E_{0}$,
owing to the commutativity of the endomorphism actions.

\subsection*{Rejection Sampling~\cite{peng2020csiibs}}
To avoid unintended disclosure of the secret key, rejection sampling can be employed. Consider two parties, $P$ and $Q$. The party $P$ select a secret touple $\mathbf{a} = (a_1, \ldots, a_n) \in [-I, I]^n$, where $I \in \mathbb{N}$. The bound $I$ is selected such that the product $\prod_{t=1}^{n} J_t^{a_t}$ spans (almost) the entire ideal class group, thereby ensuring that the resulting output distribution is close to uniform. In addition, P chooses another random vector $\textbf{a}^{\prime}$ $\in$ [-($\delta$ + 1)$I$ , ($\delta$ + 1)$I$]$^n$ where $\delta \in \mathbb{N}$. After that, Q randomly selects a bit $b$ and forwards it to P. In what follows, P computes ${\gamma} = \textbf{a}^{\prime} - \textbf{a}$ if ${b}=1$ and otherwise, $\gamma=\textbf{a}^{\prime}$. As a consequence, we can filter $\gamma$ as $|\gamma|\leq \delta I$ to prevent any compromise of the secret key. Here the output vector $\gamma$ is uniformly distributed, and as a result, it is independent of \textbf{a}.

\subsection*{Hardness Assumption}
\begin{definition}{{Group Action Inverse Problem ($\sf{GAIP}$)}~\cite{de2019seasign}:}\\
Consider two elliptic curves $E_1$ and $E_2$ over a common field with $E_2 = [\mathbf{a}]E_1$, where $[\mathbf{a}]$ belongs to the class group $\mathrm{cl}(\mathbb{Z}[\pi_{E_1}])$. Recovering $[\mathbf{a}]$ from $E_1$ and $E'$ is assumed to be a hard problem. The symbol $\pi_E$ represents the Frobenius endomorphism of $E$.
\label{GAIP}
\end{definition}
\subsection*{SeaSign~\cite{de2019seasign}}
 The construction of the SeaSign signature scheme \cite{de2019seasign} relies on three procedures, namely ${\sf Key~Generation}$, ${\sf Signing}$, and ${\sf Verification}$, which are discussed below.
\begin{enumerate}

 \item{$({PK}, {SK})\leftarrow {\sf Key~generation}(1^\lambda):$} Let $p$ be a prime and let $E_0:y^{2}=x^{3}+x$ denote a supersingular elliptic curve over $\mathbb{F}_{p}$. Taking a security parameter 
$\lambda$ as input, the signer randomly chooses $\textbf{a}$=$(a_{1},\cdots,a_{n})$ $\in$ $[-I_{0},I_{0}]^n$ and evaluates $E_{1}=[\textbf{a}]E_{0}$. The public key-secret key pair of the signer is $({PK, SK})=(E_{1}, \textbf{a})$.

 \item{$(\sigma)\leftarrow{\sf Signature}(m, {SK}, {PK}):$} On input of a message $m$, the signer performs the following steps:
    \begin{itemize}
        \item chooses  $\textbf{x}^{i}\in_{R}[-(I_{0}+I_{1}),(I_{0}+I_{1})]^n$ and evaluates $X_{i}=[\textbf{x}^{i}]E_0$ for $i=1,2,\ldots,t$,
        \item calculates $H(X_1, \cdots, X_t, m)=r_1\parallel \cdots \parallel r_t$, where $H:\{0,1\}^*\rightarrow \{0,1\}^t$ is a cryptographic hash function providing collision resistance,
        \item Define $\textbf{z}^{k}=\textbf{x}^{k}$ when $r_k=0$ ,and $\textbf{z}^{k}=\textbf{x}^{k}-\textbf{a}$ in all other cases by ensuring the fact that $\textbf{z}^{k}\in[-I_{1},I_{1}]^n$,
        \item The resulting signature is given by $$\sigma=(z_1, z_2, \cdots, z_t, r_1, r_2, \cdots, r_t).$$
    \end{itemize}
 
 \item{$(1~\mbox{or}~ 0)\leftarrow{\sf Verification}(m, \sigma, {PK}):$} On receiving a signature $\sigma$ from the sender, a verifier executes the following operations:
     \begin{itemize}
        \item computes \[
            X^{\prime}_k = \left.
            \begin{cases}
                [z_k]  \mathcal{E}_1, & \text{if } r_k=1\\
                [z_k]  \mathcal{E}, & \text{if } r_k=0

            \end{cases}
            \right\},
            \]

        \item evaluates $H(X^{\prime}_1, \cdots, X^{\prime}_t, m)=r^{\prime}_1\parallel \cdots \parallel r^{\prime}_t$ ,
        \item outputs 1 if the equality $r_1\parallel \cdots \parallel r_t= r^{\prime}_1\parallel \cdots \parallel r^{\prime}_t$ holds, else outputs 0.
    \end{itemize}
\end{enumerate}

\begin{theorem}{\cite{de2019seasign}}
The above signature scheme achieves existential unforgeability under chosen-message attacks (\emph{uf-cma}) in the random oracle model, assuming the computational hardness of the $\sf{GAIP}$ problem.

\end{theorem}

\subsection*{Formal Description of a Proxy Signature Scheme~\cite{boldyreva2012secure}}
A proxy signature scheme typically includes two participants, namely \textit{A} (the original signer) and \textit{B} (the proxy signer), with the following six algorithms:
\begin{enumerate}

  \item {{(pp)} $\leftarrow$ {${\sf Setup}$ {$(1^{\lambda})$}}:-} 
  For a chosen security parameter $\lambda\in \mathbb{N}$, this procedure determines and returns a corresponding set of system-wide public parameters, denoted by pp, which are published in a public platform like a bulletin board.

  \item {${ ((PK_A, SK_A), (PK_B, SK_B)  )}$ $\leftarrow$ ${\sf Key~generation}$ (pp):-} In this algorithm,  Both entities \textit{A} and \textit{B} execute the key generation process to obtain their own public and secret keys, represented by the pairs $(PK_{A}, SK_{A})$ and $(PK_{B}, SK_{B})$.

  \item {(${z_B}$) $\leftarrow$ ${\sf Proxy~share~generation}$ ${(SK_A,d_B)}$:-} Let $d_{B}$ be the attribute of \textit{B} consisting of the necessary information like DOB, id card number, etc. Then \textit{A} generates a proxy share $z_{B}$ with the help of $SK_{A}$ in favor of B and publishes $(d_B, z_{B})$ as a token in the bulletin board.

  \item {(0 or 1) $\leftarrow$ ${\sf Proxy~share~verification}$ $(PK_A, d_B, z_B)$:-} The entity \textit{B} executes this procedure to verify the correctness of the token $(d_B, z_{B})$. The algorithm returns 
1 if the token is deemed valid; otherwise, it outputs 0, indicating rejection.

  \item {($\sigma$) $\leftarrow$ ${\sf Proxy~signature}$ $(m,SK_B, d_B,z_B)$:-} The proxy signer \textit{B} holding a valid token, outputs a signature $\sigma$ for some message with the help $SK_B$.

   \item{(0 or 1) $\leftarrow$ ${\sf Proxy~signature~verification}$ $(m,\sigma, PK_A, PK_B, d_B, z_B)$:-} Given a message-signature pair $(m,\sigma)$ along with a token $(d_B, z_{B})$ of the proxy signer B, the verifier checks the validity of the token $(d_B, z_{B})$. If the token is valid, the verifier checks the correctness of $(m,\sigma)$ using $PK_B$. It then outputs a deterministic value of 1 if the verification is done correctly; else, it outputs 0.
\end{enumerate}

\paragraph{Correctness.} 
For any security parameter $\lambda \in \mathbb{N}$, let 
$(pp) \leftarrow {\sf Setup}(1^{\lambda})$ and 
$((PK_A, SK_A), (PK_B, SK_B)) \leftarrow {\sf Key~generation}(pp)$. 
Let $z_B \leftarrow {\sf Proxy~share~generation}(SK_A, d_B)$ such that 
$(1) \leftarrow {\sf Proxy~share~verification}(PK_A, d_B, z_B)$. 

Then, for any message $m$, it holds that
\[
\Pr\Big[
(1) \leftarrow {\sf Proxy~signature~verification}(m, \sigma, PK_A, PK_B, d_B, z_B)
\Big] = 1,
\]
where $(\sigma) \leftarrow {\sf Proxy~signature}(m, SK_B, d_B, z_B)$, and the probability is taken over the randomness of the algorithms.

\subsubsection*{Existential Unforgeability under Chosen-message Attack (UF-CMA)}
A proxy signature scheme can be specified through a set of algorithms, namely {\sf Setup}, {\sf Key Generation}, {\sf Proxy Share Generation}, {\sf Proxy Share Verification}, {\sf Proxy Signature}, and {\sf Proxy Signature Verification}. The {\em uf-cma} security for the proxy signature scheme is formalized through an interactive game between a challenger ($\mathcal{C}$) and an adversary ($\mathcal{A}$) for a particular proxy signer with attribute $d$. Once $d$ is chosen, it remains unchanged throughout the game. In this game, $\mathcal{C}$ generates the key pair for original signer and the proxy signer. In the following, $\mathcal{C}$ responds to the queries made by $\mathcal{A}$ over some messages polynomial number of times with $d^\prime$. Finally, to win the game, $\mathcal{A}$ must produce a valid message–signature pair $(m^{\star}, \sigma^{\star})$, where $m^{\star}$ is a fresh message that has not been queried earlier. We denote this experiment by $Exp^{uf-cma}_{proxy(1^{\lambda})}$, which is described below in detail.
\begin{enumerate}

    \item \textbf{Setup:} The challenger $\mathcal{C}$, using the public parameters pp along with the security parameter $\lambda$, invokes the Key Generation procedure to create two distinct key pairs: one corresponding to the original signer $(PK_{O}, SK_{O})$ and another corresponding to the proxy signer $(PK_{P}, SK_{P})$.

    \item \textbf{Sign-query:} In this phase, $\mathcal{A}$ submits a query to $\mathcal{C}$ requesting a proxy signature on a message $m$ associated with the proxy signer’s attribute $d$. $\mathcal{C}$ first recognizes the token $(d, z)$ for $d$ from the bulletin board  and executes the algorithm {\sf Proxy signature} to produce a proxy signature $\sigma$ which is sent to $\mathcal{A}$. $\mathcal{A}$ can query polynomial number of times.

    \item \textbf{Forgery:} $\mathcal{A}$ generates a forgery $\sigma^{\star}$ on a fresh message $m^{\star}$, which was not part of any prior query. $\mathcal{A}$ succeeds if it can impersonate the proxy signer i.e., {\sf Proxy signature verification} outputs 1 on input $(m^{\star},\sigma^{\star}, PK_O, PK_P, d, z)$.
\end{enumerate}
The advantage of $\mathcal{A}$ is given by $Adv_{\mathcal{A}}^{Exp^{uf-cma}_{proxy(1^{\lambda})}}$ = Prob[$Exp^{uf-cma}_{proxy(1^{\lambda})}$ =  1], i.e., the probability that $\mathcal{A}$ succeeds in the experiment.
\begin{definition}
A proxy signature scheme achieves {\em uf-cma} security if, for every probabilistic polynomial-time adversary $\mathcal{A}$, its advantage in the corresponding experiment satisfies
\begin{equation*}
        Adv_{\mathcal{A}}^{Exp^{uf-cma}_{proxy(1^{\lambda})}} \le \epsilon(\lambda)
    \end{equation*},
with the security parameter $\lambda$, $\epsilon(\lambda)$ denotes a function which is negligible.
\end{definition}

\section*{Proposed Isogeny Based Proxy Signature}\label{Proposed}

In this section, we describe our proposed scheme \textit{CSI-PS} in an Internet of Things (IoT) environment, where the IoT authority (A) acts as the original signer and delegates its signing authority to an IoT gateway (B), which acts as the proxy signer. The gateway generates proxy signatures for sensor data, while the verifier validates the authenticity of the received data. The proposed scheme consists of six algorithms: {\sf Setup}, {\sf key generation}, {\sf Proxy share generation}, {\sf Proxy share verification}, {\sf Proxy signature}, {\sf Proxy signature verification}. During the Setup phase, the IoT authority generates the public parameters required for the system. The Key Generation algorithm allows both the IoT authority (A) and the IoT gateway (B) to generate their respective public key and secret key pairs. In the Proxy Share Generation phase, the IoT authority generates a proxy share $z_{B}$ for the gateway using its secret key and the gateway’s attributes $d_B$, which may include its identity, device identifier, network address, or other relevant system information. After receiving the proxy share $z_{B}$ from the IoT authority, the gateway verifies the validity of the delegation token $(d_B, z_B)$ using the Proxy Share Verification algorithm. In the Proxy Signature phase, the gateway generates a proxy signature on the sensor data message $m$ using its secret key and the proxy share $z_{B}$. Finally, the Proxy Signature Verification algorithm is executed by the verifier (e.g., a cloud server) to check the correctness of the delegation token and the validity of the message–signature pair during the {\sf Proxy signature verification} phase. The overall workflow of the proposed CSI-PS scheme in an IoT environment is illustrated in Fig.~\ref{fig2}. The detailed description of each phase is presented below.

\begin{enumerate}

    \item {{(pp)} $\leftarrow$ {${\sf Setup}$ {$(1^{\lambda})$}}:-}On receiving the security parameter $1^\lambda$, original signer A selects suitable CSIDH (see \ref{CSIDH})  parameters including one basic elliptic curve $E_{0}:y^2=x^3 +x$, one exponent interval $[-I_{0}, I_0]$, $I_0 \in$ $\mathbb{N}$, and parameters $\alpha_{0}$, $\gamma_{0}$, $\gamma_{1}$, $M_1$, $M_2$  from $\mathbb{Z}$ .
    Then A computes one exponent interval $I_{1}=\alpha_{0}I_{0}$, two branch integers $L_{0}=2^{\gamma_{0}}-1$, $L_{1}=2^{\gamma_{1}}-1$  and publishes $pp = (E_{0}, \alpha_{0},\gamma_{0}, \gamma_{1}, M_1, M_2, I_{0}, I_{1}, L_{0}, L_{1})$
    
    \item {${ ((PK_A,SK_A), (PK_B,SK_B)  )}$ $\leftarrow$ ${\sf Key~generation}$ (pp):-}
    A does the following operations for $i=1,2,\cdots,L_{0}$:
   \begin{itemize}
       \item Randomly chooses $\textbf{a}^{(i)}$=$(a_{1}^{(i)},\cdots,a_{n}^{(i)})$ $\in$         $[-I_{0},I_{0}]^n$.
       \item Computes $E_{i}=[\textbf{a}^{(i)}]  E_{0}.$
       \item Publishes the public key $PK_{A}={(E_i)}^{L_{0}}_{i=1}$ and keeps the secret key $SK_{A}={(\textbf{a}^{(i)})}^{L_{0}}_{i=1}.$
   \end{itemize}

      On the other hand, B executes the following for $i=1,2,\cdots,L_{0}$:
    \begin{itemize}
        \item  Randomly selects $\textbf{b}^{(i)}$=$(b_{1}^{(i)},\cdots,b_{n}^{(i)})$ $\in$         $[-I_{0},I_{0}]^n$.
        \item Evaluates $\hat{E_{i}}=[\textbf{b}^{(i)}] E_{0}$.
        \item Sets the public key as $PK_{B}={(\hat{E_{i}})}^{L_{0}}_{i=1}$ and the secret key as $SK_{B}={(\textbf{b}^{(i)})}^{L_{0}}_{i=1}$.
    \end{itemize}


\begin{algorithm} \caption{Key generation (original signer)}
\begin{algorithmic}[1]
\Require pp
\ForEach{$i\in 1, \cdots, L_0$}
\State $\textbf{a}^{(i)}$=$(a_{1}^{(i)},\cdots,a_{n}^{(i)})$ $\in_R$         $[-I_{0},I_{0}]^n$.
\State Evaluate $E_{i}=[\textbf{a}^{(i)}]  E_{0}$. 
\EndFor
\Ensure $PK_{A}={(E_i)}^{L_{0}}_{i=1}$, $SK_{A}={(\textbf{a}^{(i)})}^{L_{0}}_{i=1}$.
\end{algorithmic}
\label{alg:KeyGen Original}
\end{algorithm}

\begin{algorithm} \caption{Key generation (proxy signer)}
\begin{algorithmic}[1]
\Require pp
\ForEach{$i\in 1,2, \cdots, L_0$}
\State $\textbf{b}^{(i)}$=$(b_{1}^{(i)},\cdots,b_{n}^{(i)})$ $\in_R$         $[-I_{0},I_{0}]^n$.
\State Evaluate $\hat{E_{i}}=[\textbf{b}^{(i)}]  E_{0}$. 
\EndFor
\Ensure $PK_{B}={(\hat{E_{i})}}^{L_{0}}_{i=1}$, $SK_{B}={(\textbf{b}^{(i)})}^{L_{0}}_{i=1}$.
\end{algorithmic}
\label{alg:KeyGen proxy}
\end{algorithm}

    \item {(${z_B}$) $\leftarrow$ ${\sf Proxy~share~generation}$ ${(SK_A,d_B)}$:-}
    To generate the proxy share $z_{B}$, the original signer $A$ carries out the following steps using its secret key $SK_A$ and the attribute $d_B$ of $B$. To do that, A performs the following tasks.
     \begin{itemize}
     \item Computes $h=H_{1} (d_{B},pp)$, where $H_{1} : \{{0,1}\}^{*} \rightarrow (\mathbb{Z}_{{L_0}+1})^{M_1}$ is a collision-resistant cryptographic hash function  and parses $h$ as integer $\{h_{i}\}_{i=1}^{M_1}$.
         \item For $i=1,\cdots,M_{1}$ and $j=1,\cdots,M_2$ chooses $\textbf{x}^{(i,j)}\in_{R}[-(I_{0}+I_{1}),(I_{0}+I_{1})]^n$ and evaluates, $X_{ij}=[\textbf{x}^{(i,j)}] E_{h_{i}}$.
         \item Determines $C=H_{2} (d_{B}, pp, \{{X}_{ij}\}_{i=1 j=1}^{M_{1} M_2})$ for cryptographically secure collision resistant hash function $H_{2} : \{{0,1}\}^{*} \rightarrow (\mathbb{Z}_{{L_1}+1})^{M_1 M_2}$ and parses it to integers $\{C_{(i,j)}\}_{i=1 j=1}^{M_{1} M_2}$.
         \item For $i=1,\cdots,M_{1}$ and $j=1,\cdots,M_2$, sets \\
         $\textbf{y}^{(i,j)}=\begin{cases}
                              \textbf{x}^{(i,j)}, & \text{if $C_{(i,j)} = 0$} \\
                              \textbf{x}^{(i,j)}+\textbf{a}^{(h_{i})}, & \text{otherwise}
		                      \end{cases}$
         \item If $ \textbf{y}^{(i,j)}\in [-I_{1},I_{1}]^n$ then proceeds to the next step; otherwise, it returns to step 2.
         \item Outputs the proxy share as
    $z_{B}= (C, \{\mathbf{y}^{(i,j)}\}^{M_1 , M_2}_{i=1 j=1})$ which is delivered to B and publishes $(d_B,z_B)$ as token on the bulletin board.
   \end{itemize}

\begin{algorithm} \caption{Proxy share generation}
\begin{algorithmic}[1]
\Require ${(SK_A,d_B)}$, $d_B$ is some attributes of $B$.
\State Computes $h=H_{1} (d_{B},pp)$. $H_{1} : \{{0,1}\}^{*} \rightarrow (\mathbb{Z}_{{L_0}+1})^{M_1}$ is a hash function.
\State Parses $h$ as integer $\{h_{i}\}_{i=1}^{M_1}$.

\ForEach{$i=1,\cdots,M_{1}$ and $j=1,\cdots,M_2$}
\State Chooses $\textbf{x}^{(i,j)}\in_{R}[-(I_{0}+I_{1}),(I_{0}+I_{1})]^n$.
\State Evaluates $X_{ij}=[\textbf{x}^{(i,j)}] E_{h_{i}}$.
\EndFor
\State $C=H_{2} (d_{B}, pp, \{{X}_{ij}\}_{i=1 j=1}^{M_{1} M_2})$ for a cryptographically secure collision resistant hash function $H_{2} : \{{0,1}\}^{*} \rightarrow (\mathbb{Z}_{{L_1}+1})^{M_1 M_2}$.
\State Parses $C$ to a integer set $\{C_{(i,j)}\}_{i=1 j=1}^{M_{1} M_2}$.
\ForEach{$i=1,\cdots,M_{1}$ and $j=1,\cdots,M_2$}
\If{$C_{(i,j)} = 0$}
\State $\textbf{y}^{(i,j)}=\textbf{x}^{(i,j)}$
\Else
\State $\textbf{y}^{(i,j)}=\textbf{x}^{(i,j)}+\textbf{a}^{(h_{i})}$
\EndIf
\If{$ \textbf{y}^{(i,j)}\in [-I_{1},I_{1}]^n$}
\State Perform the next step.
\Else
\State Start from Step 3.
\EndIf
\EndFor
\Ensure $z_{B}= (C, \{\mathbf{y}^{(i,j)}\}^{M_1 , M_2}_{i=1 j=1})$ and publishes $(d_B,z_B)$ as the token.
\end{algorithmic}
\label{alg:Proxyshare gen}
\end{algorithm}

    \item {(0 or 1) $\leftarrow$ ${\sf Proxy~share~verification}$ $(PK_A, d_B, z_B)$:-}
    With $(PK_A, d_B, z_B)$ provided, the proxy signer B performs the following tasks to check whether the token $(d_B,z_B)$ is valid or not:
    \begin{itemize}
        \item Determines $h=H_{1}(d_{B},pp)$, and parses $h$ and $C$ into integers as $\{h_{i}\}_{i=1}^{M_1}$ with $0\leq h_{i}\leq L_{0}$ and $\{C_{(i,j)}\}_{i=1 j=1}^{M_{1} M_2}$ with  $0\leq C_{(i,j)} \leq L_1$ respectively.
        \item For $i=1,\cdots,M_{1}$ and $j=1,\cdots,M_2$ sets\\
        $\overline{X}_{ij}=\begin{cases}
                              [\textbf{y}^{(i,j)}] E_{h_{i}}, & \text{if $C_{(i,j)} = 0$} \\
                              [\textbf{y}^{(i,j)}] E_{0}, & \text{otherwise}
		                      \end{cases}$

       \item Determines $C^{\prime}=H_{2}(d_{B}, pp, \{\overline{X}_{ij}\}_{i=1 j=1}^{M_{1} M_2})$.
        \item If $C=C^{\prime}$, outputs 1 and gets the assurance of signing authority on behalf of the original signer A; otherwise, outputs 0.
    \end{itemize}

\begin{algorithm} \caption{Proxy share verification}
\begin{algorithmic}[1]
\Require $(PK_A, d_B, z_B)$.
\State Evaluate $h=H_{1} (d_{B},pp)$.
\State Parses $h$ as integers $\{h_{i}\}_{i=1}^{M_1}$, $0\leq h_{i}\leq L_{0}$.
\State Parses $C$ as integers $\{C_{(i,j)}\}_{i=1 j=1}^{M_{1} M_2}$, $0\leq C_{(i,j)} \leq L_1$.
\ForEach{$i=1,\cdots,M_{1}$ and $j=1,\cdots,M_2$}
\If{$C_{(i,j)} = 0$}
\State $\overline{X}_{ij}=[\textbf{y}^{(i,j)}] E_{h_{i}}$
\Else
\State $\overline{X}_{ij}=[\textbf{y}^{(i,j)}] E_{0}$
\EndIf
\State $C^{\prime}=H_{2}(d_{B}, pp, \{\overline{X}_{ij}\}_{i=1 j=1}^{M_{1} M_2})$.
\If{$C=C^{\prime}$}
\State True.
\Else
\State False.
\EndIf
\EndFor
\Ensure True/False.
\end{algorithmic}
\label{alg:Proxyshare Ver}
\end{algorithm}

    \item {($\sigma$) $\leftarrow$ ${\sf Proxy~signature}$ $(m,SK_B, d_B,z_B)$:-}
    On input of a message $m \in \{0,1\}^{\star}$, $SK_{B}$, $d_B$ and $z_{B}$, B  executes the following steps to generate a proxy signature $\sigma$ for $m$.Given a message $m \in \{0,1\}^{\star}$ along with the inputs $SK_{B}$, $d_B$ and $z_{B}$, the proxy signer B carries out the following procedure to produce a proxy signature $\sigma$ corresponding to $m$.
    \begin{itemize}
    \item Computes $h^{\prime}={H_1}(z_{B},pp)$ and parses $h^{\prime}$ as integers $\{h_{i}^{\prime}\}_{i=1}^{M_1}$ for $0\leq h_{i}^{\prime}\leq L_{0}$.
        \item For $i=1,\cdots,M_{1}$ and $j=1,\cdots, M_2$, chooses  $\boldsymbol{\xi}^{(i,j)}\in_{R}[-(I_{0}+I_{1}),(I_{0}+I_{1})]^n$ and evaluates $Y_{(i,j)}=[\boldsymbol{\xi}^{(i,j)}]{\hat{E}_{h_{i}^{\prime}}}$.
        \item Determines $D={H_2}(z_{B}, pp, m, \{{Y}_{ij}\}_{i=1 j=1}^{M_{1} M_2})$ and parses it to integers $\{D_{(i,j)}\}_{i=1 j=1}^{M_{1} M_2}$ with  $0\leq D_{(i,j)} \leq L_1$.
        \item For $i=1,\cdots,M_{1}$ and $j=1,\cdots,M_2$ sets\\
         $\boldsymbol{\eta}^{(i,j)}=\begin{cases}
                              \boldsymbol{\xi}^{(i,j)}, & \text{if $D_{(i,j)} = 0$} \\
                              \boldsymbol{\xi}^{(i,j)}+ \textbf{b}^{(h^{\prime}_i)}, & \text{otherwise}
		                      \end{cases}$
      \item If $\boldsymbol{\eta}^{(i,j)}\in [-I_{1},I_{1}]^n$ after which the process continues to the next step; otherwise, it restarts from Step 2.
      \item Outputs the signature as
   ${\sigma=\{D, {\{\boldsymbol{\eta}}^{(i,j)}\}_{i=1 j=1}^{M_{1} M_{2}}}\}$.
    \end{itemize}
    
\begin{algorithm} \caption{Proxy signature}
\begin{algorithmic}[1]
\Require $(m,SK_B, d_B,z_B)$, $m\in \{0,1\}^{\star}$
\State Computes $h^{\prime}={H_1}(z_{B},pp)$.
\State $h^{\prime}=\{h_{i}^{\prime}\}_{i=1}^{M_1}$, $h_{i}^{\prime}$'s are integers with $0\leq h_{i}^{\prime}\leq L_{0}$.

\ForEach{$i=1,\cdots,M_{1}$ and $j=1,\cdots,M_2$}
\State Chooses $\boldsymbol{\xi}^{(i,j)}\in_{R}[-(I_{0}+I_{1}),(I_{0}+I_{1})]^n$.
\State Evaluates $Y_{(i,j)}=[\boldsymbol{\xi}^{(i,j)}]{\hat{E}_{h_{i}^{\prime}}}$.
\EndFor
\State $D={H_2}(z_{B}, pp, m, \{{Y}_{ij}\}_{i=1 j=1}^{M_{1} M_2})$.
\State Parses $D$ to a integer set $\{D_{(i,j)}\}_{i=1 j=1}^{M_{1} M_2}$ with  $0\leq D_{(i,j)} \leq L_1$.
\ForEach{$i=1,2,\ldots,M_{1}$ and $j=1,2,\ldots,M_2$}
\If{$D_{(i,j)} = 0$}
\State $\boldsymbol{\eta}^{(i,j)}=\boldsymbol{\xi}^{(i,j)}$
\Else
\State $\boldsymbol{\eta}^{(i,j)}= \boldsymbol{\xi}^{(i,j)}+ \textbf{b}^{(h^{\prime}_i)}$
\EndIf
\If{$\boldsymbol{\eta}^{(i,j)}\in [-I_{1},I_{1}]^n$}
\State Perform the next step.
\Else
\State Start from Step 3.
\EndIf
\EndFor
\Ensure ${\sigma=\{D, {\{\boldsymbol{\eta}}^{(i,j)}\}_{i=1 j=1}^{M_{1} M_{2}}}\}$.
\end{algorithmic}
\label{alg:Proxysig gen}
\end{algorithm}

   \item{(0 or 1) $\leftarrow$ ${\sf Proxy~signature~verification}$ $(m,\sigma, PK_A, PK_B, d_B, z_B)$:-}
   Using $(PK_A, d_B, z_B)$, the verifier initially checks the validity of the token $(d_B,z_B)$ of B using a similar approach as discussed in {\sf proxy share verification}. If the verification fails it outputs $0$; otherwise, does the following:
   \begin{itemize}
       \item Computes $h^{\prime}={H_1}(z_{B},pp)$, and parses $h^{\prime}$ and $D$ into integers as $\{h_{i}^{\prime}\}_{i=1}^{M_1}$ with $0\leq h_{i}^{\prime}\leq L_{0}$ and $\{D_{(i,j)}\}_{i=1 j=1}^{M_{1} M_{2}}$ with  $0\leq D_{(i,j)}\leq L_1$, respectively.
       \item For $i=1,\cdots,M_{1}$ and $j=1,\cdots,M_2$ , sets\\
       $\overline{Y}_{(i,j)}=\begin{cases}
                              [{\boldsymbol{\eta}}^{(i,j)}]{\hat{E}_{h_{i}^{\prime}}}, & \text{if $D_{(i,j)} = 0$} \\
                              [{\boldsymbol{\eta}}^{(i,j)}]{{E}_{0}}, & \text{otherwise}
		                      \end{cases}$
       \item Evaluates $D^{\prime}={H_2}(z_{B}, pp, m, \{{\overline{Y}}_{ij}\}_{i=1 j=1}^{M_{1} M_{2}})$.
   \item  If $D=D^{\prime}$, outputs 1; otherwise, outputs 0.

\begin{algorithm}[h!] \caption{Proxy signature verification}
\begin{algorithmic}[1]
\Require $(m,\sigma, PK_A, PK_B, d_B, z_B)$.
\State Evaluate $h=H_{1} (d_{B},pp)$.
\State Parses $h$ as integers $\{h_{i}\}_{i=1}^{M_1}$, $0\leq h_{i}\leq L_{0}$.
\State Parses $C$ as integers $\{C_{(i,j)}\}_{i=1 j=1}^{M_{1} M_2}$, $0\leq C_{(i,j)} \leq L_1$.
\ForEach{$i=1,\cdots,M_{1}$ and $j=1,\cdots,M_2$}
\If{$C_{(i,j)} = 0$}
\State $\overline{X}_{ij}=[\textbf{y}^{(i,j)}] E_{h_{i}}$
\Else
\State $\overline{X}_{ij}=[\textbf{y}^{(i,j)}] E_{0}$
\EndIf
\State $C^{\prime}=H_{2}(d_{B}, pp, \{\overline{X}_{ij}\}_{i=1 j=1}^{M_{1} M_2})$.
\If{$C=C^{\prime}$}
\State Determines $h^{\prime}={H_1}(z_{B},pp)$.
\State Parses $h^{\prime}$ as $\{h_{i}^{\prime}\}_{i=1}^{M_1}$, where $h_{i}^{\prime}$'s are integers with $0\leq h_{i}^{\prime}\leq L_{0}$.
\State Parses $D$ as integers $\{D_{(i,j)}\}_{i=1 j=1}^{M_{1} M_{2}}$ with  $0\leq D_{(i,j)}\leq L_1$.
\ForEach{$i=1,\cdots,M_{1}$ and $j=1,\cdots,M_2$}
\If{$D_{(i,j)} = 0$}
\State $\overline{Y}_{(i,j)}=[{\boldsymbol{\eta}}^{(i,j)}]{\hat{E}_{h_{i}{\prime}}}$
\Else
\State $\overline{Y}_{(i,j)}=[{\boldsymbol{\eta}}^{(i,j)}]{{E}_{0}}$
\EndIf
\State $D^{\prime}={H_2}(z_{B}, pp, m, \{{\overline{Y}}_{ij}\}_{i=1 j=1}^{M_{1} M_{2}})$.
\If{$D=D^{\prime}$}
\State True.
\Else
\State False.
\EndIf
\EndFor
\Else
\State False
\EndIf
\EndFor
\Ensure True/False.
\end{algorithmic}
\label{alg:Proxysig Ver}
\end{algorithm}
   
   \end{itemize}
\end{enumerate}

\begin{figure}[htbp]
\centering
\includegraphics[width=0.4\textwidth]{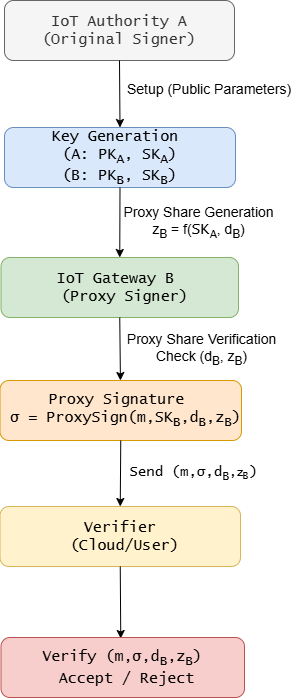}
\caption{Workflow of the proposed CSI-PS scheme in an IoT environment.}
\label{fig2}
\end{figure}
	
   \section*{Security}\label{Security}
   In this section, we analyze the security properties of the proposed scheme and examine its resistance against potential attacks.
  \begin{itemize}

      \item \textbf{Unforgeability of proxy signature:} 
      The {\em uf-cma} security of the \textit{CSI-PS} scheme is examined under the assumption that the isogeny-based $\sf{GAIP}$P problem is computationally hard. The hash functions $H_1$ and $H_2$ are assumed to behave as random oracles. Our proof proceeds via contradiction by assuming the existence of an adversary $\mathcal{A}$ that achieves a non-negligible advantage in the {\em uf-cma} security game. Then we are going to show that using $\mathcal{A}$, an oracle machine $O^{\mathcal{A}}$ can be designed to break the isogeny problem by utilizing the outputs of $H_1$ and $H_2$. To prove this claim, we define a collection of games \textit{$Game^0$}, \textit{$Game^1$}, \textit{$Game^2$}, and \textit{$Game^3$}, such that for each $i = 1,2,3$, \textit{$Game^i$} differs from \textit{$Game^{i-1}$} only by a small modification. Let us denote the success probability of $\mathcal{A}$ in $Game^i$ as Prob[{$Game^i$}].
      \begin{enumerate}
          \item {$\textbf{Game}^0$:} \textit{$Game^0$} is equivalent to the UF-CMA security game for proxy signatures. Therefore, 
          $Adv_{\mathcal{A}}^{Exp^{uf-cma}_{proxy(1^{\lambda})}}$ = Prob[$Exp^{uf-cma}_{proxy(1^{\lambda})}$ =  1] = Prob[$Game^0$].
          \item {$\textbf{Game}^1$}: This game is similar to the previous {$Game^0$}, other from that $O^{\mathcal{A}}$ replaces output of the the hash function $H_1$ on $(z_{B},pp)$ by a random string of length $M_1$ from $(\mathbb{Z}_{{L_0}+1})^{M_1}$. If $|Prob(Game^1)-Prob(Game^0)|$ is non-negligible, then $\mathcal{A}$ will be able to distinguish the output distribution of $H_1$, which is impossible since $H_1$ is chosen as a random oracle. Hence  $|Prob(Game^1)-Prob(Game^0)|$ should be negligible, say bounded by $\epsilon_{1}(\lambda)$.
          \item{$\textbf{Game}^2$}: {$Game^2$} is similar to {$Game^1$}, except that $O^{\mathcal{A}}$ substitutes output of the hash function $H_2$ by a random string chosen from $(\mathbb{Z}_{{L_1}+1})^{M_1 M_2}$. By a similar argument as discussed in $Game^1$, $|Prob(Game^2)-Prob(Game^1)|$ $\le$ $\epsilon_{2}(\lambda)$ (negligible).
          \item {$\textbf{Game}^3$}: This game is similar to the previous  \textbf{$Game^2$}, other from that in the challenge stage when $\mathcal{A}$ tries to forge a signature for a message $m^{*}$, $O^{\mathcal{A}}$ substitutes $H_1$ output by a random tuple from $(\mathbb{Z}_{{L_0}+1})^{M_1}$ and substitute $H_2$  by a random tuple from $(\mathbb{Z}_{{L_1}+1})^{M_1 M_2}$. Using the similar argument as discussed in \textit{$Game^1$}, we may conclude that $|Prob(Game^3)-Prob(Game^2)|$ $\le$  $\epsilon_{3}(\lambda)$ (negligible).
      \end{enumerate}
      Now
      $|Prob(Game^3)-Adv_{\mathcal{A}}^{Exp^{uf-cma}_{\textit{CSI-PS}(1^{\lambda})}}|$\\
      $=|Prob(Game^3)-Prob(Game^0)|$\\
      $\leq$ $|Prob(Game^3)-Prob(Game^2)|$+$|Prob(Game^2)-Prob(Game^1)| +|Prob(Game^1)-Prob(Game^0)|$\\
      $\le$ $\epsilon_{1}(\lambda)$+$\epsilon_{2}(\lambda)$+$\epsilon_{3}(\lambda)=\epsilon(\lambda)$ (say).\par
      Therefore, in $Game^3$, the probability that $\mathcal{A}$ succeeds is approximately $Adv_{AD}^{Exp^{uf\text{-}cma}_{\textit{CSI-PS}(1^{\lambda})}}$. Since we have assumed that $Adv_{\mathcal{A}}^{Exp^{uf-cma}_{\textit{CSI-PS}(1^{\lambda})}}$ is non-negligible we can ensure that $Prob(Game^3)$ is also non-negligible.
      Thereby, with the help of $\mathcal{A}$ and controlling $H_1$, $H_2$, the oracle machine \textit{$O^{\mathcal{A}}$} can generate two valid transcripts $(\{{Y}_{ij}\}$, $D^{(1)}$, ${\boldsymbol{\eta}_{1}^{(i,j)}})$ and $(\{{Y}_{ij}\}, D^{(2)}, {\boldsymbol{\eta}_{2}^{(i,j)}})$  where $D^{(1)}_{(i,j)}=0$ and $D^{(2)}_{(i,j)} \neq 0$ $\forall$ $i$, $j$. Note that $\boldsymbol{\eta}_{1}^{(i,j)}=\boldsymbol{\xi}^{(i,j)}$ since $D^{(1)}_{(i,j)}=0$ and $\boldsymbol{\eta}_{2}^{(i,j)}=\boldsymbol{\xi}^{(i,j)}+\textbf{b}^{\left(h^{\prime}_i \right)}$ since $D^{(2)}_{(i,j)} \neq 0$. Utilizing $\boldsymbol{\eta}_{1}^{(i,j)}$ and $\boldsymbol{\eta}_{2}^{(i,j)}$, \textit{$O^{\mathcal{A}}$} can easily extract the secret $SK_B = \{B^i\}_{i=1}^{L_0}$. Therefore, the ($\sf{GAIP}$) problem between two elliptic curves (see Def. \ref{GAIP}) is broken, which leads to a contradiction. Thus, the advantage of $\mathcal{A}$ in the \textit{uf-cma} game can not be non-negligible. In other words, our scheme \textit{CSI-PS} is uf-cma secure. \\
      \indent Next, we discuss the remaining security properties of the proposed \textit{CSI-PS} scheme.

\item \textbf{Identifiability:} Given a signature $\sigma$ with the proxy share $z_B$, the verifier finds $(d_B, z_B)$ from the bulletin board and verifies the correctness of $z_B$. If the verification succeeds, the identity of the proxy signer can be determined from $d_B$.
   \item \textbf{Undeniability:} During the verification phase, the proxy signer cannot deny having generated the proxy signature. This is because the proxy signature includes the components ${{\mathbf{\eta}^{(i,j)}}{i=1, j=1}^{M{1} L_{1}}}$, which are constructed using the proxy signer’s secret key. Therefore, once the proxy signature is produced, the proxy signer cannot repudiate their involvement in its creation.
   \item \textbf{Verifiability:} During verification, the verifier needs to confirm that the proxy signature comes from a legitimate delegation established between the original signer and the proxy signer. The token $z_B$ is created using the secret key of original signer’s. By checking whether $z_B$ is valid with respect to $PK_A$, the verifier becomes confident that such an agreement exists.
   
   \item \textbf{Distinguishability:} A verifier who has permission to validate signatures can clearly differentiate between a proxy signature and a regular signature. In an isogeny-based scheme such as SeaSign, a standard signature is produced solely with the signer’s private key and is authenticated with the corresponding public key. In contrast, within a proxy signature framework, the proxy signer generates the signature by using both their own private key and a proxy component received from the original signer. As part of the verification process, the verifier utilizes this proxy component along with the public details of both the original signer and the proxy signer. Due to these differences in the signing and verification procedures, signatures produced via proxy delegation are distinguishable from standard signatures.
   \item \textbf{Secrecy:} The proxy share $z_B$ remains independent of the original signer’s secret key in terms of information leakage. Recovering this key from the token would require solving instances of the $\sf{GAIP}$ problem (see Definition \ref{GAIP}) for pairs $(X_{ij}, E_{h_i})$, where $i = 1, \ldots, M_1$ and $j = 1, \ldots, M_2$. Since this problem is computationally hard, the secret key of the original signer cannot be feasibly recovered in practice. A similar argument applies to the proxy signer, whose secret key cannot be efficiently derived using a proxy signature.
   \item \textbf{Prevention of misuse:} The proxy signer can sign a message using the token $(d_B, z_B)$ assigned specifically to them. This token includes their identity and other necessary information. Therefore, any misuse of the proxy signature can be traced back to the proxy signer through the token.
   \item \textbf{Revocability:} The proxy signer's signature authority is immediately revoked after the token $(d_B,z_B)$'s time period expires. Besides, the IoT authority can broadcast the revocation message of invalidation of the token $(d_B,z_{B})$ by putting a signed message on the bulletin board.
  \end{itemize}

\section*{Efficiency}
\label{Sec: efficiency}

The proxy signature scheme proposed by Mathieu de Goyon and Atsuko Miyaji~\cite{de2024isogeny} is mainly focused on the cryptographic construction of a proxy signature using the CSI-FiSh~\cite{beullens2019csi} framework. Their work presents a theoretical design of an isogeny-based proxy signature and analyzes its security properties. However, the scheme does not consider a specific application environment or system architecture where such a delegation mechanism can be practically deployed.

In contrast, the proposed scheme extends the use of isogeny-based proxy signatures to the Internet of Things (IoT) environment. Our work introduces an IoT-oriented delegation framework in which an authority can delegate signing rights to proxy nodes such as gateways that generate authenticated signatures for sensor data. This approach provides a practical mechanism for secure authentication and scalable data verification in resource-constrained IoT systems, thereby demonstrating how post-quantum proxy signatures can be effectively integrated into real-world IoT infrastructures.
  
  \subsection*{Complexity Analysis of {\it CSI-PS}}\label{complexity}
 We describe below the computational complexity and communication cost of our proposed \textit{CSI-PS}. The overall communication cost is influenced by the sizes of the public keys (original and proxy), the proxy share, and the proxy signature. Each signer's public key is made up of $L_0$ number of elliptic curves. According to Theorem \ref{Montgomery}, any elliptic curve can be encoded as a single coefficient over $\mathbb{F}_p$ after class group action. Additionally, proxy share generation and proxy signature generation both contain one hash output and $M_1M_2$ vectors from $[-I_1, I_1]^n$. In our scheme, we employ two hash functions, denoted by $H_i$ for $i\in\{1,2\}$. In proxy share generation and proxy signature generation algorithms, four hash outputs $h, h^{\prime}$, $C$, and $D$ have been used. The lengths of $h$ and $h^{\prime}$ are $M_{1}\left \lceil{log(L_{0}+1)}\right \rceil$ bits. The lengths of $C$ and $D$ are $M_{1} M_{2}\left \lceil{log(L_{1}+1)}\right \rceil$ bits. The communication complexity of our proposed scheme \textit{CSI-PS} is given in Table~\ref{keysize}.\\
 To ensure $\lambda$ bits of security, we may adopt $M_{1}\left \lceil{log(L_{0}+1)}\right \rceil \geq \lambda$ and $M_{1} M_{2}\left \lceil{log(L_{1}+1)}\right \rceil \geq \lambda$. Since $L_{0} = 2^{\gamma_{0}}-1$ and $L_{1} = 2^{\gamma_{1}}-1$, we conclude that $M_{1}{\gamma_{0}} \geq \lambda$ and $M_{1} M_{2}{\gamma_{1}} \geq \lambda$. In particular, we select our own parameters $M_1$ and $M_2$ as
  $M_1 \gamma_{0} =\lambda $, $M_1 M_2\gamma_{1}=\lambda$. Hence, $M_1 = \frac{\lambda}{\gamma_{0}}$, $M_1 M_2 = \frac{\lambda}{\gamma_{1}}$ and $M_2 = \frac{\gamma_{0}}{\gamma_{1}}$. We choose $\alpha_0=nM_1 L_1$. Castryck et al. \cite{castryck2018csidh} proposed parameters for CSIDH-512 to achieve AES-128 bit security. CSIDH-512 includes the prime of the type $p=2^2l_{1}l_{2}\ldots l_{n}-1$ with, $n=74$, $l_{73} = 373$ and $l_{74}=587$, the key bound $m$ is taken as 5 (See subsection \ref{CSIDH} for details of the parameters). In our scheme $n=74$, $I_0 = 5$. So as to calculate the sizes of the keys, proxy share and proxy signature, we rely on the different choices of parameters $(\gamma_{0}, \gamma_{1}, M_1, M_2, L_{0}, L_{1})$ as proposed in \cite{peng2020csiibs} for $\lambda = 128$-bit security level. For some choices of parameters the results are documented in Table~\ref{Communication}.

\begin{table}[h]
 	\caption{The sizes of public keys, secret keys, proxy share, and proxy signature.}
 	\centering
  \renewcommand{\arraystretch}{1.5}
    \begin{tabular}{||p{4cm} | p{7cm}||}

 \hline
 Data & Size (bits) \\ [0.5ex]
 \hline\hline
 $PK_A$ &  $L_{0}\left \lceil{log(p)}\right \rceil $\\
 \hline
 $SK_A$ &  $L_{0}\left \lceil{n log(2I_{0}+1)}\right \rceil $\\
 \hline
 $PK_B$ &  $L_{0}\left \lceil{log(p)}\right \rceil $ \\
 \hline
 $SK_B$ &  $L_{0}\left \lceil{n log(2I_{0}+1)}\right \rceil $ \\
 \hline
 Proxy share($z_B$) &  $M_{1}M_{2}\left \lceil{log(L_{1}+1)}\right \rceil +$  $M_{1}M_{2}\left \lceil{n log(2I_{1}+1)}\right \rceil $\\
 \hline
 Proxy Signature($\sigma$) & $M_{1}M_{2}\left \lceil{log(L_{1}+1)}\right \rceil +$ $M_{1}M_{2}\left \lceil{n log(2I_{1}+1)}\right \rceil $ \\[1ex]
 \hline
\end{tabular}
\label{keysize}
 \end{table}

 \begin{table}
 	\caption{Communication cost and secret key size for different parameters for 128 bit security parameters}
 	\centering
   \renewcommand{\arraystretch}{1.5}
    \begin{tabular}{{||p{4cm}|p{3cm}|p{3cm}|p{3cm}||}}

 \hline
 $(\gamma_{0}, \gamma_{1}, M_1, M_2, L_{0}, L_{1})$ & Public key size & Secret key size & Proxy share or signature size\\ [0.5ex]
 \hline\hline
(1, 1, 128, 1, 1, 1) & 63.75B & 32B & 19584B \\
 \hline
(1, 8, 16, 1, 1, 255) & 63.75B & 32B & 3198B\\
\hline
(8, 16, 16, 1, 255, 65535) & 16256.25B & 8160B & 4402B\\
\hline
(16, 8, 8, 2, 65535, 255) & 4177856.25B & 2097120B & 3052B\\
\hline
\end{tabular}
\label{Communication}
 \end{table}

 \subsection*{Concrete Instantiation in IoT}\label{toy}
This section describes a practical implementation of the proposed isogeny-based proxy signature technique in an IoT environment for a 128-bit security level with the parameter set (16, 8, 8, 2, 65535, 255). Consider an IoT network where an IoT authority (original signer) is responsible for managing and authenticating the information collected from multiple IoT devices. However, because of the numerous connected devices and continuous data generation, it may not be efficient for the authority to directly sign every message. Therefore, the authority delegates its signing rights to an IoT gateway (proxy signer), which acts on behalf of the authority to carry out signing processes. In this scenario, IoT sensors collect environmental data and transmit it to the gateway node. The gateway then generates proxy signatures for the received data and forwards the signed information to the cloud server for verification. We demonstrate the delegation process and key generation using a toy example.

\begin{itemize}
    \item Here $I_0 = 5$, $n=74$, $\gamma_{0}=16$, $\gamma_{1}=8$, $M_1=8$, $M_2=2$, $L_{0}=2^{16} -1 = 65535$, $L_{1} = 2^8 -1 = 255$, $I_1=754800$
\end{itemize}
\begin{itemize}
  \item A selects the vectors $\textbf{a}^1, \cdots, \textbf{a}^{65535}$ randomly from $[-5,5]^{74}$.
    \item A computes $E_1, \cdots, E_{65535}$, where $E_i=[\textbf{a}^{(i)}]E_0$, $i=1, \cdots, 65535$.
\end{itemize}
\begin{itemize}
  \item B selects the vectors $\textbf{b}^1, \cdots, \textbf{b}^{65535}$ randomly from $[-5,5]^{74}$.
    \item B computes $\hat{E_{1}},\cdots, \hat{E}_{65535}$, where $\hat{E_{i}}=[\textbf{b}^{(i)}]E_0$, $i=1, \cdots, 65535$.
\end{itemize}
$SK_A=\textbf{a}^1, \cdots, \textbf{a}^{65535}$, $PK_A=E_1, \cdots, E_{65535}$ and $SK_B = \textbf{b}^1, \cdots, \textbf{b}^{65535}$, $PK_B = \hat{E_{1}}, \cdots, \hat{E}_{65535}$.
The original signer will now perform the following actions.
\begin{itemize}
    
    \item $h = H_1(d_B,pp)={h_1, \cdots, h_8}$, where $h_1, \cdots, h_8 \in \mathbb{Z}_{65536}$.
    \item Randomly selects, $\textbf{x}^{(1,1)}, \textbf{x}^{(1,2)}, \textbf{x}^{(2,1)}, \textbf{x}^{(2,2)}, \cdots, \textbf{x}^{(8,1)},$ $ \textbf{x}^{(8,2)}$ from $[754795,754805]^{74}$
    \item Evaluates $X_{11}=[\textbf{x}^{(1,1)}]E_{h_{1}}$, $X_{12}=[\textbf{x}^{(1,2)}]E_{h_{1}}$, $X_{21}=[\textbf{x}^{(2,1)}]E_{h_{2}}$, $X_{22}=[\textbf{x}^{(2,2)}]E_{h_{2}}$, $\cdots$, $X_{81}=[\textbf{x}^{(8,1)}]E_{h_{8}}$, $X_{ij}=[\textbf{x}^{(8,2)}]E_{h_{8}}$.
    \item $C=H_{2} (d_{B}, pp, \{{X}_{ij}\}_{i=1 j=1}^{8~~2})$= $\{C_{(1,1)}, C_{(1,2)}, C_{(2,1)}, C_{(2,2)},\cdots, C_{(8,1)}, C_{(8,2)}\}\subset \mathbb{Z}_{256}$.
    \item Computes $\textbf{y}^{(1,1)}, \textbf{y}^{(1,2)}, \textbf{y}^{(2,1)}, \textbf{y}^{(2,2)}, \cdots, \textbf{y}^{(8,1)}, \textbf{y}^{(8,2)}$ according to Algorithm \ref{alg:Proxyshare gen}.
\end{itemize}
Therefore the proxy share is $z_{B}= (C, \{\mathbf{y}^{(i,j)}\}^{8~~2}_{i=1 j=1})$. Using Algorithm \ref{alg:Proxyshare Ver}, the representative proxy signer can validate the proxy share using its own credentials. In the following, the representative signer B may be granted the original signer's signing rights. B can now use the delegated signing rights to generate proxy signatures on IoT sensor data on behalf of the IoT authority A. We now demonstrate how the IoT gateway B can produce a proxy signature for the sensed data message.

\begin{itemize}
    \item Evaluates $h^{\prime}={H_1}(z_{B},pp)=\{h_{1}^{\prime}, \cdots, h_{8}^{\prime}\}$, where  $h_{1}^{\prime}, \cdots, h_{8}^{\prime}\in \mathbb{Z}_{65536}$.
    \item Randomly selects, $\boldsymbol{\xi}^{(1,1)}, \boldsymbol{\xi}^{(1,2)}, \boldsymbol{\xi}^{(2,1)}, \boldsymbol{\xi}^{(2,2)}, \cdots, \boldsymbol{\xi}^{(8,1)},$ $\boldsymbol{\xi}^{(8,2)}$ from $[754795,754805]^{74}$.
    \item Computes the following elliptic curves, $Y_{11}=[\textbf{x}^{(1,1)}]\hat{E}_{h_{1}^{\prime}}$, $Y_{12}=[\boldsymbol{\xi}^{(1,2)}]\hat{E}_{h_{1}^{\prime}}$, $Y_{21}=[\boldsymbol{\xi}^{(2,1)}]\hat{E}_{h_{2}^{\prime}}$, $Y_{22}=[\boldsymbol{\xi}^{(2,2)}]\hat{E}_{h_{2}^{\prime}}$, $\cdots$, $Y_{81}=[\boldsymbol{\xi}^{(8,1)}]\hat{E}_{h_{8}^{\prime}}$, $Y_{ij}=[\boldsymbol{\xi}^{(8,2)}]\hat{E}_{h_{8}^{\prime}}$.
    \item Assume that $m$ represents the sensed data message generated by an IoT device that needs to be authenticated. To proceed further he calculates the hash value, $D=H_{2} (z_{B}, pp, m,  \{{Y}_{ij}\}_{i=1 j=1}^{8~~2})$= $\{D_{(1,1)}, D_{(1,2)}, D_{(2,1)}, D_{(2,2)},\cdots, D_{(8,1)}, D_{(8,2)}\}\subset \mathbb{Z}_{256}$.
    \item Computes $\boldsymbol{\eta}^{(1,1)}, \boldsymbol{\eta}^{(1,2)}, \boldsymbol{\eta}^{(2,1)}, \boldsymbol{\eta}^{(2,2)}, \cdots, \boldsymbol{\eta}^{(8,1)}, \boldsymbol{\eta}^{(8,2)}$ by the method given in Algorithm \ref{alg:Proxysig gen} and checks whether $\boldsymbol{\eta}^{(1,1)}, \boldsymbol{\eta}^{(1,2)}, \boldsymbol{\eta}^{(2,1)}, \boldsymbol{\eta}^{(2,2)}, \cdots, \boldsymbol{\eta}^{(8,1)}, \boldsymbol{\eta}^{(8,2)}\in [-754800,754800]^{74}$.
    
\end{itemize}
When the procedure is complete, the proxy signer produces the proxy signature $\sigma={D, {{\boldsymbol{\eta}}^{(i,j)}}_{i=1, j=1}^{8~~2}}$. The cloud server, acting as the verifier, validates this signature using Algorithm \ref{alg:Proxysig Ver}.\\
\indent We now discuss the computation complexity of {\it CSI-PS} for $128$-bit security level corresponding to aforementioned parameter set. Within \textit{CSI-PS}, class group computation represents the primary computational bottleneck. In the first step, we estimated the cost of one class group computation expressed using field multiplication operations. In the following, we compute the computational complexity of public key generation, proxy share generation, and proxy signature generation in terms of total number of field multiplications required. We also calculated the time complexity of {\it CSI-PS} by estimating the run time and cpu clock cycles for public key generation, proxy share generation, and proxy signature generation. The implementation was carried out in C and executed on a workstation equipped with an Intel Core i5-8700 2.40,GHz processor, 8,GB RAM, and Linux Lite v5.2 OS. The experimental results are summarized in Table~\ref{Compuntationtion}.

\begin{table}
 \caption{Computation complexity and time complexity of {\it CSI-PS} for 128 bit security level}
 \centering
 { \renewcommand{\arraystretch}{1.5}
\begin{tabular}{||p{4cm}|p{3cm}|p{3cm}|p{3cm}||}
 \hline& Clock cycle & Time (in seconds) & Field multiplications\\\hline
CSIDH-512 (one class group computation) & $45.7 \times 10^6$ & $0.021624$ & $282,424$ \\\hline
 Public Key &  $29.94 \times 10^{11}$ & $1422.12$ & $18, 508, 656, \newline 840$ \\ \hline
Proxy Share & $73.12 \times 10^8$ & $0.3472$  & $4, 518, 784$ \\\hline
Proxy Signature& $73.12 \times 10^8$ & $0.3472$  & $4, 518, 784$ \\\hline
\end{tabular}}
\label{Compuntationtion}
 \end{table}

\section*{Conclusion}\label{Conclusion}
We develop a framework that allows the delegation of signing capabilities from an original signer to proxy signers by leveraging isogeny-based cryptographic constructions. The isogeny-based signature scheme {\em SeaSign} \cite{de2019seasign} serves as the fundamental building block for our scheme. Our proposed scheme achieves {\em uf-cma} security under the assumption that the ($\sf{GAIP}$) problem between two elliptic curves is hard. Additionally, our construction meets the usual security guarantees associated with proxy signatures. In our design, the proxy signature is $M_1M_2\left \lceil{n \log(2I_{1}+1)}\right \rceil$ + $M_1M_2\left \lceil{\log(L_{1}+1)}\right \rceil $  bits in size. Through calculations, we conclude that the proxy signature size is 3052B for a 128-bit security level with a certain parameter choice. Constructing the proxy signature requires 16 class group operations or 4,518,784 field multiplications. 

In addition, the proposed CSI-PS scheme is well-suited for Internet of Things (IoT) environments, where resource-constrained devices benefit from efficient delegation of signing operations to more capable nodes such as gateways. This significantly reduces computational overhead at sensor devices while maintaining secure authentication and data integrity. Future work includes optimizing computational efficiency and exploring real-world deployment in large-scale IoT infrastructures.

\section*{Acknowledgments}
The paper was edited for grammar using ``Grammarly'' (\url{https://www.grammarly.com/}).

\section*{Declarations}
\subsection*{Ethical approval:} The research does not involve any Human Participants and/or Animals.
\subsection*{Conflicts of interest:} There are no potential conflicts of interest among the authors.
\subsection*{Data availability:} The datasets used and/or analysed during the current study are available from the corresponding author on reasonable request.
\subsection*{Funding:} This work was supported by the Abu Dhabi University's Office of Research and Sponsored Programs under Grant 19300906.
\subsection*{Author contributions statement}
``SK - Conceptualization, Investigation, Supervision, Resources, Formal analysis, Writing -- original draft, Writing -- review \& editing KD - Conceptualization, Formal analysis, Resources, Validation, Writing -- original draft, Writing -- review \& editing
VS- Conceptualization, Validation, Investigation, Writing -- original draft, Writing -- review \& editing
SKD- Conceptualization, Supervision, Validation, Investigation, Writing -- review \& editing
AKD- Supervision, Validation, Investigation, Writing -- review \& editing
SAC- Supervision, Validation, Investigation, Funding, Writing -- review \& editing''

\bibliography{ref}

\end{document}